\documentclass[%
 reprint,
superscriptaddress,
 amsmath,amssymb,
 aps,
prd,
]{revtex4-1}

\usepackage{graphicx}
\usepackage{dcolumn}
\usepackage{bm}

\begin{document}


\title{Cosmic Transients, Einstein's Equivalence Principle and  Dark Matter Halos}

\author{Orfeu Bertolami}\email{orfeu.bertolami@fc.up.pt }
 \affiliation{Departamento de F\'isica e Astronomia,\\
Faculdade de Ci\^encias da Universidade do Porto,\\
Rua do Campo Alegre 687, 4169-007, Porto, Portugal}
 \author{Ricardo G. Landim}\email{rlandim@if.usp.br}
 \affiliation{Departamento de F\'isica e Astronomia,\\
Faculdade de Ci\^encias da Universidade do Porto,\\
Rua do Campo Alegre 687, 4169-007, Porto, Portugal}
\affiliation{%
Instituto de F\'isica, Universidade de S\~ao Paulo\\
 Rua do Mat\~ao,
1371, Butant\~a, CEP 05508-090, S\~ao Paulo, SP, Brazil
}

\date{\today}

\begin{abstract}
Cosmic transients, such as gamma-ray bursts and fast radio bursts, have been used to constrain the Einstein's Equivalence Principle (EEP) trough the parametrized-post-Newtonian (PPN) formalism. In this approach, the time delay of photons with different energies from these cosmic transients  are used to obtain   upper bounds on the difference of the PPN $\gamma$ parameter. In this work we assume that an important contribution to the time delay is due to  the dark matter halo of the Milky Way and  consider the dark matter mass distribution given by the Navarro--Frenk--White profile. We obtain  the upper limit on the difference of the PPN parameter $\gamma$ for the polarized gamma-ray emission of GRB 110721A,  $\Delta \gamma < 1.06 \times 10^{-28}$, the most stringent limit to date on the EEP. In addition,  we show that a very similar upper bound is obtained if, instead of having a dark matter component, a visible matter density profile and a non-minimal gravitational coupling between curvature and matter are present.

\end{abstract}

\maketitle

\section{\label{sec:level1}Introduction} 
General relativity (GR) has been tested  through several and repeated experiments and observations, since a few years after the appearance of the final version of the Einstein field
equations. Although the success of the theory   to provide a comprehensive understanding of several  gravitational phenomena, the lack of knowledge regarding the nature of dark matter and dark energy encourages the search for deviations from GR at astrophysical and cosmological scales.  One broadly used framework to compare Einstein's theory with other viable metric theories of gravity is  the parametrized post-Newtonian (PPN) formalism \cite{eddington1923the,robertson1962relativity,schiff1962general,1967rta1.book..105S,nordtvedt1968equivalence,nordtvedt1968equivalence,will1971theoretical,nordtvedt1972conservation} (see also Refs. \cite{Will:2014kxa,bertolami2014experimental} for   reviews on the experimental status of GR). Making use of the PPN parameters, such as $\gamma$, the Einstein's Equivalence Principle (EEP) can be checked and constrained \cite{Will:2014kxa,bertolami2014experimental,Gao:2015lca}. 

Taking either two different particles (e.g. photons and neutrinos) or two (same) particles with  different energies into account (therefore with different $\gamma_1$ and $\gamma_2$), it is possible to constrain EEP since the theories of gravity that are built in agreement with it predict $\gamma_1=\gamma_2$. The accuracy of the EEP has been tested through the difference of the $\gamma$ values, $\Delta \gamma \equiv \gamma_1-\gamma_2$, using several astronomical sources: supernova 1987A \cite{Longo:1987gc,Krauss:1987me},  gamma-ray bursts (GRBs) \cite{Gao:2015lca,Wei:2016ygk,Sang:2016egz,Wu:2017yjl,Yang:2017pww}, fast radio bursts (FRBs) \cite{Wei:2015hwd,Tingay:2016tgf,Nusser:2016wzr,Wu:2017yjl}, the Crab pulsar \cite{Yang:2016cfi,Desai:2016nqu,Zhang:2016myf}, blazars \cite{Wei:2016mzj,Wang:2016lne}
and gravitational wave (GW)  sources \cite{Wu:2016igi,Kahya:2016prx,Li:2016zjc,Wei:2017nyl}. Very recently, the association of GW and GRB observations was used  to provide a new constraint on the $\Delta \gamma$ through the time delay between the gravitational and electromagnetic radiation produced by a binary neutron star merger \cite{Monitor:2017mdv}. The most stringent  established bound, $\Delta \gamma <1.6\times 10^{-27}$, was obtained    using the measurement of polarized gamma-ray photons from GRB  \cite{Yang:2017pww}.

Many of the previous studies (including the most stringent limit on EEP  \cite{Yang:2017pww}) assumed that the time delays are caused mainly by the gravitational potential of the Milky
Way, in which only the visible matter plays a role. However, the contribution of the dark matter halo may have an influence on the EEP limits, since it increases the total mass of our galaxy. 

N-body simulations can provide reliable theoretical predictions for halos surrounding galaxies such as the Milky Way. One of the most used profiles to describe  the mass distribution of several galaxies, including our own, is the Navarro--Frenk--White (NFW) profile \cite{Navarro:1996gj}.

On the other hand, modified gravity theories attempt to explain the  effect of the dark matter through a different set of assumptions (see Ref. \cite{Clifton:2011jh} for a review). Among several possibilities, it is viable to mimic  known dark matter density profiles (among them, the NFW profile) through a non-minimal gravitational coupling between   matter and curvature \cite{Bertolami:2009ic}. In this approach, the NFW profile can be mimicked by visible matter and a power-law coupling.

Independent of what gives origin to the dark matter profile, i.e., the self-assembling of dark matter particles into a halo  or the mimicking mechanism, the presence of such mass distribution does affect the previous EEP limits.

Therefore, in this letter, we  investigate the influence of the dark matter halo of the Milky Way on the EEP accuracy, where we assume that the density distribution of the halo follows the NFW profile. We also analyze the changes in these limits with the mimicking mechanism. In order to constrain $\Delta \gamma$ and compare the results,  we consider observations from one polarized GRB \cite{Yang:2017pww},  three FRBs \cite{Wei:2015hwd} and the recently observed association of   GW and GRB.  \cite{Monitor:2017mdv}.

This letter is organized in the following manner. In section \ref{eepfrb} we review the procedure of testing EEP with cosmic transients. In section \ref{sec:nfw} we use the NFW profile to obtain  upper bounds on EEP. Section \ref{mimicking} is devoted for the mimicking dark matter approach. We summarize our results in section \ref{summary}.

\section{Constraining EEP with cosmic transients}\label{eepfrb}

The observed time delay for two particles with different energies from cosmic transient sources has  the following contributions
\cite{Gao:2015lca,Wei:2015hwd}:
\begin{equation}\label{tobs}
\Delta t_{obs}=\Delta t_{int}+\Delta t_{LIV}+\Delta t_{spe}+\Delta t_{DM}+\Delta t_{grav}\, ,
\end{equation}
where $\Delta t_{int}$ is the intrinsic astrophysical time delay between two test photons, $\Delta t_{LIV}$ represents the time delay due to Lorentz invariance violation, $\Delta t_{spe}$ is  the time delay caused by  photons with a rest mass different from zero, $\Delta t_{DM}$ is the time delay from the dispersion by the line-of-sight free electron content and, finally, $\Delta t_{grav}$ is the  time delay of two photons of energy $E_1$ and $E_2$ due to the gravitational potential $\Phi(r)$ integrated from the source to the Earth. 

The gravitational time delay $\Delta t_{grav}$ yields
\begin{equation}\label{tgrav}
\Delta t_{grav}=\frac{\Delta \gamma}{c^3}\int_{r_0}^{r_e}\Phi(r)dr\, ,
\end{equation}
where  $\gamma$ is the PPN parameter, $r_0$  and $r_e$ are 	the positions of the observer and  the source, respectively. 

Hereinafter both $\Delta t_{LIV}$ and $\Delta t_{spe}$ will be, as discussed in Refs. \cite{bertolami2014experimental,Gao:2015lca}, neglected. Assuming that $\Delta t_{int}>0$ and that $\Delta t_{obs}\gg \Delta t_{DM} $ \footnote{The effect of the dispersion process on $\Delta \gamma$ can be seen  in Fig. 1 of Ref. \cite{Wei:2015hwd}, where a range from $\Delta t_{DM} = 0.999\Delta t_{obs}$ to $\Delta t_{DM} = 0.001\Delta t_{obs}$ was analyzed. For the purposes of this work we take the same limit $\Delta t_{obs}\gg \Delta t_{DM} $.}, from Eqs. (\ref{tobs}) and (\ref{tgrav}) we have
\begin{equation}\label{deltagamma}
\Delta \gamma<c^3\Delta t_{obs} \left(\int_{r_0}^{r_e}\Phi(r)dr\right)^{-1}\, .
\end{equation}

The first constraint on EEP using FRBs observations was performed in Ref. \cite{Wei:2015hwd}, where the authors assumed the Keplerian potential and obtained upper limits for $\Delta \gamma $ using three observations: FRB 110220  \cite{Thornton:2013iua} and associations of FRB with GRB 101011A \cite{Cannizzo2010} and with GRB 100704A \cite{McBreen2010,Grupe2010}. In these FRB/GRB associations   \cite{bannister2012limits}, the redshift was inferred using the Amati relation \cite{Deng:2013aga,Amati:2002ny}. 

The most stringent limit on EEP in the literature was also obtained using the Keplerian potential. The difference   $\Delta\gamma$ was obtained between the two orientations of a circularly polarized gamma-ray with an energy of 70 keV  \cite{Yang:2017pww}. The value $\Delta \gamma <1.6\times 10^{-27}$ was reached using the measurement of those polarized gamma-ray photons \cite{Yonetoku:2012wz} of the GRB 110721A \cite{2011GCN.11771....1F,2011GCN.12187....1T}. The resulted time lag from the rotation of
the linear polarization angle is $\Delta t_{grav}=\Delta \phi\lambda/(2 \pi c) $, where $\lambda$ is the wavelength and the upper limit for $\Delta \phi $ is $2\pi$. 

The time delay of $+1.74$ s between the GW170817 and GRB 170817A was used to place upper and lower bounds on the difference between the speed of gravity and the speed of light. Taking the limits of the time delay, $-8.26$ s $\leq \Delta t \leq +1.74$ s, the lower bound of the 90\% confidence level on the luminosity distance derived from the GW signal ($D=26 $ Mpc)   \cite{TheLIGOScientific:2017qsa} and the effect of the Keplerian potential outside a sphere of $100 $ kpc, the constraint on $\Delta \gamma$ is $-2.6 \times 10^{-7}\leq \gamma_{GW}-\gamma_{EM}\leq 1.2 \times 10^{-6}$ \cite{Monitor:2017mdv}.  

The inferred redshifts of the sources, as well as the arrival time delay $\Delta t_{obs}$  and the correspondent  upper limits for $\Delta \gamma$, from Refs. \cite{Wei:2015hwd,Yang:2017pww}, are shown in Table \ref{tableEEP}. For comparison purposes, we use the  observations from Table \ref{tableEEP} and from Ref. \cite{Monitor:2017mdv} in order to constrain $\Delta \gamma$ assuming a dark matter halo described by the NFW profile. 

 Notice that there are intermediate upper bounds on $\Delta \gamma$ in the literature, obtained, for instance,  through photon emission from the Crab pulsar ($\Delta \gamma < 10^{-15}$) \cite{Yang:2016cfi} or  polarized photons from a FRB, whose   main contribution in the gravitational potential arises from the Laniakea supercluster of galaxies ($\Delta \gamma < 10^{-16}$) \cite{Wu:2017yjl}.  However, these are not suitable to access the influence of the dark matter on the EEP accuracy. The Crab pulsar is in  the Milky Way, where the  dark matter contribution is sub-dominant. The task of finding a dark matter profile for clusters of galaxies is much more involved than for galaxies, and thus the comparison among the EEP limits using the gravitational potential of the Milky Way and the NFW profile is  more evident.

\begin{table}[h]
\begin{ruledtabular}
\begin{tabular}{ c c c c}
	 & $z$ & $\Delta t_{obs}$ (s) & $\Delta \gamma$\\
\hline
FRB 110220  &  0.81  	  &  1     &  $ 2.52\times 10^{-8} $ \cite{Wei:2015hwd}\\
 FRB/GRB 101011A &  0.246 &  0.438  & $ 1.24 \times 10^{-8}$ \cite{Wei:2015hwd}\\
 FRB/GRB 100704A  &    0.166    &     0.149     & $ 4.36 \times 10^{-9}$  \cite{Wei:2015hwd}\\
  Polarized GRB 110721A &  0.382 & $6\times 10^{-20}$ & $ 1.6\times 10^{-27}$ \cite{Yang:2017pww}\\
\end{tabular}
\end{ruledtabular}
\caption{Upper limits of $\Delta \gamma$, for three different FRB observations \cite{Wei:2015hwd} and one GRB observation \cite{Yang:2017pww}.  }\label{tableEEP}
\end{table}

\section{Dark matter mass distribution}\label{sec:nfw}

The NFW profile is given by \cite{Navarro:1996gj} 
\begin{equation}\label{nfwprofile}
\rho_{\textrm{NFW}}(r)=\frac{\rho_s}{(r/r_s)\left(1+r/r_s\right)^2}\, ,
\end{equation}
where $\rho_s$ is a reference scale density, defined at a scale radius $r=r_s$.

Following the notation of Ref. \cite{Wang:2015ala}, the NFW potential is found from Eq. (\ref{nfwprofile}) through the integration of the Poisson equation \cite{Navarro:1995iw,Navarro:1996gj}:
\begin{widetext}
\begin{align}
\Phi(r)=&-4 \pi G \rho_s r_s^2\left(\frac{\ln (1+r/r_s)}{r/r_s}+\frac{1}{1+r_{\text{max}}/r_s}\right)\quad \text{for} \quad  r<r_{\text{max}}\,,\label{phi1}\\
\Phi(r)=&-4 \pi G \rho_s r_s^2\left(\frac{\ln (1+r_{\text{max}}/r_s)}{r/r_s}+\frac{r_{\text{max}}/r_s}{(r/r_s)(1+r_{\text{max}}/r_s)}\right) \quad  \text{for} \quad r> r_{\text{max}} \,,\label{phi2}
\end{align}
\end{widetext}
where $r_{max}$ is the halo boundary. As we shall see, the halo boundary does not affect significantly our estimates. 

The scale density is related with the concentration parameter $c_{200}\equiv r_{200}/r_s$, where $r_{200}$ is the virial radius, through the relation \cite{Navarro:1995iw}:
\begin{equation}\label{c200}
\rho_{s}=\rho_c \left(\frac{200}{3}\right) \frac{c_{200}^3}{\ln(1 + c_{200}) - c_{200}/(1 + c_{200})}\, ,
\end{equation}
where $\rho_c$ is the critical density of the Universe, $\rho_c=1.879$ $ h^2 \times 10^{-29} $g cm$^3$ and $h=0.7$.

We substitute Eqs. (\ref{phi1})--(\ref{c200}) into Eq. (\ref{deltagamma}) to constrain $\Delta \gamma$. In order to test the influence of the halo boundary we choose different values of $r_{max}$, expressed in terms of the virial radius, and we  also considered the limit when the boundary is infinite ($r_{max} \rightarrow \infty$). In this limit only the first term in Eq. (\ref{phi1}) contributes. 

The two independent parameters in the NFW potential, namely, $c_{200}$ and $r_s$, depend on the structure of the dark matter halo. On the other hand, N-body simulations provide   predictions for halos surrounding galaxies such as the Milky Way. These parameters can be obtained, for instance, by the Aquarius  \cite{Springel:2008cc}  and the Via Lactea \cite{Diemand:2008in} simulations.  Here we take  five halos  from the Aquarius N-body simulation in the context of the standard $\Lambda$CDM cosmology \cite{Navarro:2008kc} that is in agreement with an estimate using dynamical tracers \cite{Wang:2015ala},  to bound the EEP. The values for the scale radius and for the concentration parameter are shown in Tables \ref{results} and \ref{results2}.

The results in Tables \ref{results} and \ref{results2} show that the limits on EEP are around one order of magnitude more stringent than the ones of previous studies \cite{Yang:2017pww,Wei:2015hwd,Monitor:2017mdv}. Neither the difference between halos nor the halo boundary change significantly $\Delta \gamma$. Greater values of $r_{max}$ make the bounds on EEP slightly more stringent. Although  the halos presented in Ref. \cite{Navarro:2008kc} have different values of $c_{200}$ and $r_{s}$  than the ones of Ref. \cite{Wang:2015ala} (or even in other simulations, as in Illustris simulation \cite{Taylor:2015jaa}), these differences  do not have a major influence on our results. The most stringent limits on $\Delta \gamma $ are from the ones that use the halo D. 

\begin{table*}
\begin{ruledtabular}
\begin{tabular}{ c c c c c c c}
$r_{max} [r_{200}]$ &	 & A  & B  & C  & D   &   E\\
     \hline
& FRB 110220  &  $4.17 \times 10^{-9}$  	  &  $7.68 \times 10^{-9}$     &  $ 4.07\times 10^{-9} $  &  $3.70 \times 10^{-9}$     &  $5.47 \times 10^{-9}$    \\
2& FRB/GRB 101011A &   $2.10\times 10^{-9}$  &  $3.86\times 10^{-9}$   & $ 2.05 \times 10^{-9}$   &  $1.86 \times 10^{-9}$     & $2.75\times 10^{-9}$     \\
& FRB/GRB 100704A  &  $7.47 \times 10^{-10}$    &    $1.37\times 10^{-9}$      & $ 7.29 \times 10^{-10}$  &  $6.65 \times 10^{-10}$     & $9.80\times 10^{-10}$   \\
& Polarized GRB 110721A &  $2.28 \times 10^{-28}$    &    $4.19\times 10^{-28}$      & $2.22 \times 10^{-28}$  &  $2.02 \times 10^{-28}$     & $2.98\times 10^{-28}$   \\
\hline
& FRB 110220  &  $3.91 \times 10^{-9}$  	  &  $7.11 \times 10^{-9}$     &  $ 3.80\times 10^{-9} $  &  $3.43 \times 10^{-9}$     &  $5.07 \times 10^{-9}$    \\
3& FRB/GRB 101011A &   $1.97\times 10^{-9}$  &  $3.58\times 10^{-9}$   & $ 1.92 \times 10^{-9}$   &  $1.74 \times 10^{-9}$     & $2.56\times 10^{-9}$     \\
& FRB/GRB 100704A  &  $7.04 \times 10^{-10}$    &    $1.28\times 10^{-9}$      & $ 6.84 \times 10^{-10}$  &  $6.20 \times 10^{-10}$     & $9.13\times 10^{-10}$   \\
& Polarized GRB 110721A &  $2.13 \times 10^{-28}$    &    $3.87\times 10^{-28}$      & $2.07 \times 10^{-28}$  &  $1.87 \times 10^{-28}$     & $2.76\times 10^{-28}$   \\
\hline
& FRB 110220  &  $3.75 \times 10^{-9}$  	  &  $6.76 \times 10^{-9}$     &  $3.63\times 10^{-9} $  &  $3.27 \times 10^{-9}$     &  $4.83 \times 10^{-9}$    \\
4& FRB/GRB 101011A &   $1.90\times 10^{-9}$  &  $3.42\times 10^{-9}$   & $ 1.84 \times 10^{-9}$   &  $1.66 \times 10^{-9}$     & $2.44\times 10^{-9}$     \\
& FRB/GRB 100704A  &  $6.78 \times 10^{-10}$    &    $1.22\times 10^{-9}$      & $ 6.57 \times 10^{-10}$  &  $5.93 \times 10^{-10}$     & $8.73\times 10^{-10}$   \\
& Polarized GRB 110721A &  $2.04 \times 10^{-28}$    &    $3.68\times 10^{-28}$      & $1.97 \times 10^{-28}$  &  $1.78 \times 10^{-28}$     & $2.62\times 10^{-28}$   \\
\hline
& FRB 110220  &  $2.52 \times 10^{-9}$  	  &  $4.22\times 10^{-9}$     &  $2.39\times 10^{-9} $  &  $2.09 \times 10^{-9}$     &  $3.05 \times 10^{-9}$    \\
$\infty$& FRB/GRB 101011A &   $1.39\times 10^{-9}$  &  $2.35\times 10^{-9}$   & $ 1.33 \times 10^{-9}$   &  $1.17 \times 10^{-9}$     & $1.70\times 10^{-9}$     \\
& FRB/GRB 100704A  &  $5.13 \times 10^{-10}$    &    $8.66\times 10^{-10}$      & $ 4.89 \times 10^{-10}$  &  $4.31 \times 10^{-10}$     & $6.28\times 10^{-10}$   \\
& Polarized GRB 110721A &  $1.29 \times 10^{-28}$    &    $2.14\times 10^{-28}$      & $1.21 \times 10^{-28}$  &  $1.06 \times 10^{-28}$     & $1.55\times 10^{-28}$   \\
\end{tabular}
\end{ruledtabular}
\caption{ Upper limits for $\Delta \gamma$ using the observations in Table \ref{tableEEP}. Five different halos are shown  (A [$c_{200}= 16.10  $, $r_{s}= 15.49$ kpc], B [$c_{200}= 8.16  $, $r_{s}= 23.32$ kpc], C [$c_{200}= 12.34  $, $r_{s}= 19.96$ kpc], D [$c_{200}= 8.73  $, $r_{s}= 28.19$ kpc] and E [$c_{200}= 8.67  $, $r_{s}= 24.83$ kpc]) \cite{Navarro:2008kc}, with different halo boundaries $r_{max}$. }\label{results}
\end{table*}

\begin{table*}
\begin{ruledtabular}
\begin{tabular}{  c c c c c c}
$r_{max} [r_{200}]$ 	 & A  & B  & C  & D   &   E\\
     \hline
2&     $7.43 \times 10^{-8}$  	  &  $1.35 \times 10^{-7}$     &  $ 7.22\times 10^{-8} $  &  $6.55 \times 10^{-8}$     &  $9.65 \times 10^{-8}$    \\
&    $-1.57\times 10^{-8}$  &  $-2.85\times 10^{-8}$   & $ -1.52 \times 10^{-8}$   &  $-1.38 \times 10^{-8}$     & $-2.03\times 10^{-8}$     \\
\hline
3&  $7.06 \times 10^{-8}$  	  &  $1.27 \times 10^{-7}$     &  $ 6.84\times 10^{-8} $  &  $6.17 \times 10^{-8}$     &  $9.06 \times 10^{-8}$    \\
&    $-1.49\times 10^{-8}$  &  $-2.67\times 10^{-8}$   & $ -1.44 \times 10^{-8}$   &  $-1.30 \times 10^{-8}$     & $-1.91\times 10^{-8}$     \\
\hline
4&   $6.85 \times 10^{-8}$  	  &  $1.22 \times 10^{-7}$     &  $ 6.62\times 10^{-8} $  &  $5.96 \times 10^{-8}$     &  $8.74 \times 10^{-8}$    \\
&    $-1.44\times 10^{-8}$  &  $-2.57\times 10^{-8}$   & $ -1.39 \times 10^{-8}$   &  $-1.26 \times 10^{-8}$     & $-1.84\times 10^{-8}$     \\
\hline
$\infty$&   $6.54 \times 10^{-8}$  	  &  $1.12 \times 10^{-7}$     &  $ 6.29\times 10^{-8} $  &  $5.63 \times 10^{-8}$     &  $8.15 \times 10^{-8}$    \\
&    $-1.38\times 10^{-8}$  &  $-2.36\times 10^{-8}$   & $ -1.33 \times 10^{-8}$   &  $-1.19 \times 10^{-8}$     & $-1.72\times 10^{-8}$     \\
\end{tabular}
\end{ruledtabular}

\caption{ Upper (positive values) and lower (negative values) limits for $\Delta \gamma=\gamma_{GW}-\gamma_{EM}$ using GW170817 and GRB 170817A \cite{Monitor:2017mdv}. Five different halos are shown  (A [$c_{200}= 16.10  $, $r_{s}= 15.49$ kpc], B [$c_{200}= 8.16  $, $r_{s}= 23.32$ kpc], C [$c_{200}= 12.34  $, $r_{s}= 19.96$ kpc], D [$c_{200}= 8.73  $, $r_{s}= 28.19$ kpc] and E [$c_{200}= 8.67  $, $r_{s}= 24.83$ kpc]) \cite{Navarro:2008kc}, with different halo boundaries $r_{max}$. }\label{results2}
\end{table*}

\section{Mimicking the dark matter density profile}\label{mimicking}

It was shown in Ref. \cite{Bertolami:2009ic} that it is possible to fit the rotation curves of some  selected galaxies considering, instead of a dark matter component, a visible matter density profile and a non-minimal gravitational coupling between curvature and matter \cite{Bertolami:2007gv}. 

The model, in agreement with Solar System tests  \cite{Bertolami:2013qaa}, has the action given by
\begin{equation}\label{lagrangianMDE}
S=\int \sqrt{-g} \left[\frac{1}{16 \pi G} R+\left(\frac{R_0}{R}\right)^n\mathcal{L}_m\right]d^4x\, ,
\end{equation}
 where $R_0$ is a characteristic scale. The non-minimal coupling can  mimic known dark matter density profiles depending on the exponent $n$. NFW profile is mimicked for $n=1/3$. The mimicking dark matter approach  is  valid in the outer region, where the
curvature is  enough so that the effect of the non-minimal gravitational coupling becomes manifest. The effect of the non-minimal coupling between matter and curvature on the features and on the dispersion relation of gravitational waves has been discussed in Ref. \cite{Bertolami:2017svl}.

In the outer region, the dark matter profile would exhibit a behavior dominated by a power-law for distances above a threshold $r_s$, thus the matched NFW profile is
\begin{equation}\label{nfwouter}
\rho_{NFW,\,outer}(r)\approx \frac{\rho_s}{(r/r_s)^{3}}\, .
\end{equation}
Although in Eq. (\ref{deltagamma}) the potential should be integrated in regions that not  lie only in the outer region, it is possible to investigate how sensitive are the results presented in the last section when the NFW potential takes the approximate form given by Eq. (\ref{nfwouter}). For simplicity and since $\Delta \gamma$ is not significantly changed, we assume an infinite halo boundary. Thus the potential in this outer region is
\begin{equation}\label{nfwprofileouter}
\Phi_{outer}(r)\approx -4 \pi G \rho_s r_s^2\left(\frac{\ln (r/r_s)}{r/r_s}\right)\, .
\end{equation}

Notice that for GW170817 and GRB 170817A the NFW potential $\Phi(r)$ is integrated  from $100$ kpc  to $26$ Mpc \cite{Monitor:2017mdv}, thus the radius does lie entirely in the outer region and our results using either Eq. (\ref{nfwprofile}) or Eq. (\ref{nfwouter}) are exactly the same. 

Using the same set of observations that led to Table \ref{results}, we depict in Table \ref{resultsII} the upper limits on EEP obtained by gravity model, Eq. (\ref{lagrangianMDE}). When comparing the last three rows in Table \ref{results} (corresponding to $r_{max} \rightarrow \infty$) with the results in Table \ref{resultsII}, we see that $\Delta \gamma$ is slightly changed, indicating that the upper limits are not sensitive whether we use the full form of the NFW profile, Eq. (\ref{nfwprofile}) or the approximation in the outer region, Eq. (\ref{nfwouter}).

\begin{table*}\begin{ruledtabular}
\begin{tabular}{ c c c c c c c}
$r_{max} [r_{200}]$ &	 & A  & B  & C  & D   &   E\\
     \hline
& FRB 110220  &  $2.57 \times 10^{-9}$  	  &  $4.32\times 10^{-9}$     &  $ 2.44\times 10^{-9} $  &  $2.15 \times 10^{-9}$     &  $3.13 \times 10^{-9}$    \\
$\infty$& FRB/GRB 101011A &   $1.43\times 10^{-9}$  &  $2.42\times 10^{-9}$   & $ 1.36 \times 10^{-9}$   &  $1.21 \times 10^{-9}$     & $1.76\times 10^{-9}$     \\
& FRB/GRB 100704A  &  $5.25 \times 10^{-10}$    &    $8.97\times 10^{-10}$      & $5.04 \times 10^{-10}$  &  $4.50 \times 10^{-10}$     & $6.51\times 10^{-10}$   \\
& Polarized GRB 110721A &  $1.30 \times 10^{-28}$    &    $2.18\times 10^{-28}$      & $1.24 \times 10^{-28}$  &  $1.08 \times 10^{-28}$     & $1.58\times 10^{-28}$   \\

\end{tabular}
\end{ruledtabular}
\caption{ Upper limits for $\Delta \gamma$ in the outer region, using the  observations in Table \ref{tableEEP} and the same five different halos in Table \ref{results}. }\label{resultsII}
\end{table*}

\section{Summary}\label{summary}

The accuracy of the EEP can be tested through astrophysical transients, such as GRBs and FRBs.  Among different gravitational potentials assumed to  contribute to the time delay, the Keplerian potential of the Milky Way was used in  previous studies (see Table I from Ref. \cite{Wei:2017nyl}). By taking the dark matter halo of the Milky Way, described by the NFW profile, we have obtained  more stringent values of $\Delta \gamma$ than the corresponding previous results \cite{Wei:2015hwd,Yang:2017pww,Monitor:2017mdv}.

As can be seen when comparing Tables \ref{tableEEP} and \ref{results}, and the results from Ref. \cite{Monitor:2017mdv} with Table  \ref{results2},   the introduction of dark matter improves the previous upper bounds on  EEP by around one order of magnitude. In fact, using the polarized GRB photons from GRB 110721A we obtained the most stringent  bound to date on EEP. The halo boundary or different halo structures change slightly the results. Among the five halos analyzed, the most stringent value is obtained for the halo D with an infinite  boundary. 

Similar  results are obtained for the  mimicking dark matter mechanism, in which the NFW profile is mimicked through a power-law gravitational coupling between matter and curvature. 

Therefore, the contribution due to the NFW profile does change  the bounds on EEP. On the other hand, the limits are not accurate enough to the origin of such dark matter profile, i.e., the NFW profile can arise either from the self-assembling of dark matter particles into a halo or  from a visible matter with non-minimal coupling between matter and curvature.

\acknowledgments
RL thanks CNPq (Conselho Nacional de Desenvolvimento Cient\'ifico e Tecnol\'ogico, grant 150254/2017-2, Brazil) for the financial support and the Departamento de F\'isica e Astronomia, Faculdade de Ci\^encias, Universidade do Porto for hosting him while the work was in progress.


\providecommand{\noopsort}[1]{}\providecommand{\singleletter}[1]{#1}%

\end{document}